# Simplified and Secure MCP Gateways for Enterprise AI Integration

Ivo Brett CISSP, B.Eng, MSc
Solution Architect / Educator
independent.academia.edu/ivobrett

*Abstract—* The increasing adoption of the Model Context Protocol (MCP)[1] for AI Agents necessitates robust security for Enterprise integrations. This paper introduces the MCP Gateway to simplify secure self-hosted MCP server integration.[2] The proposed architecture integrates security principles, authentication, intrusion detection, and secure tunneling, enabling secure self-hosting without exposing infrastructure. Key contributions include a reference architecture, threat model mapping, simplified integration strategies, and open-source implementation recommendations. This work focuses on the unique challenges of enterprise-centric, self-hosted AI integrations, unlike existing public MCP server solutions.[3]

## I. Introduction

The Model Context Protocol (MCP) enhances AI systems by enabling dynamic interaction with external tools but introduces critical security risks.[4] Enterprises implementing MCP servers face amplified challenges, as developers must manage OAuth, API security, and threat mitigation—diverting focus from core AI integration tasks. To address this, we propose the MCP Gateway, a dedicated intermediary that centralizes security, monitoring, and policy enforcement. The latest versions of the MCP Specification[2] mandates OAuth 2.1 and Dynamic Client Registration, increasing compliance complexity. The Gateway abstracts these requirements, handling authentication, identity integration, and access control while ensuring spec-compliant deployments. As AI adoption grows, tailored security for protocols like MCP becomes essential. The MCP Gateway mitigates risks without overburdening developers, enabling secure, scalable AI-tool integration.

## II. Contributions

This paper makes three key contributions: (a) A reference architecture for MCP Gateways validated through implementation; (b) Threat model mappings with corresponding mitigation strategies; and (c) Tool-specific implementation guidelines.

## III. Methodology

The research methodology comprised: (1) Analysis of MCP security challenges through technical documentation and community discourse; (2) Development of security controls compliant with MCP protocol standards; and (3) Empirical validation via prototype implementation. The approach balanced standardization requirements with practical deployment considerations.

## IV. Architectural Separation of MCP Components

The evolution of MCP's security requirements has led to thoughtful enhancements that better align with enterprise security practices[8]. The 2025-03-26 MCP specification[2] introduced OAuth 2.1 support, prompting a clear distinction between the resource server (responsible for tool execution) and the authorization server (managing OAuth flows). This separation reflects a growing focus on scalability, simplified token management, and alignment with zero-trust architecture—ultimately making the specification more adaptable and robust for enterprise environments.

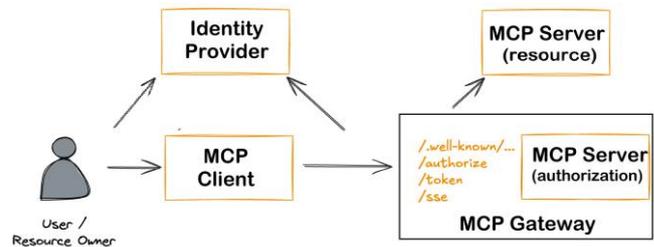

Fig. 1.  Separation of Resource and Authorization in MCP Server.

This led to the conceptual separation of concerns: the MCP Gateway assumes authorization responsibilities (OAuth 2.1 flows, token validation, identity integration) while MCP servers focus solely on resource provision. The Gateway acts as a policy enforcement point, translating enterprise identity tokens into MCP-specific credentials through a dedicated authentication service. This architecture aligns with enterprise patterns where API gateways front specialized services, while maintaining compliance with MCP specifications through protocol-level interoperability. The separation reduces attack surface (isolating sensitive auth logic) and simplifies server development - critical for adoption in security-conscious environments.

## V. Reference Architecture

The MCP Gateway provides a layered security architecture designed to protect self-hosted MCP Servers.

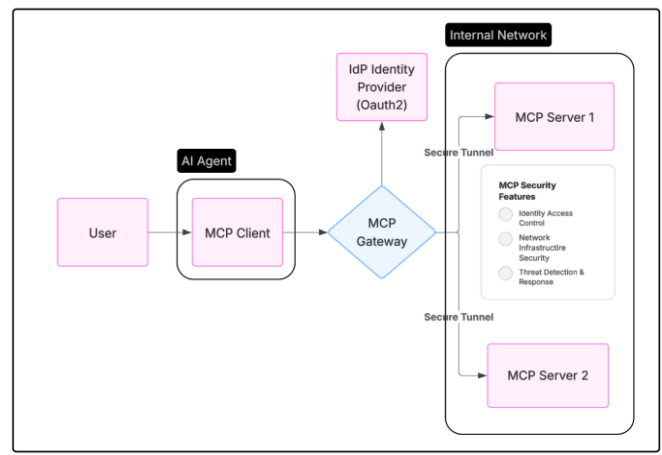

Fig. 2.  Reference Architecture

https://orcid.org/0009-0003-6373-153X

The MCP gateway's core components include:

- Security Proxy: Handles ingress traffic with TLS termination, rate limiting, and forward authentication delegation.
- Authentication Gateway: Manages OAuth 2.1 flows, integrates with enterprise identity providers, and validates tokens while offloading auth from MCP servers.
- Zero Trust Tunnelling: Establishes identity-aware encrypted tunnels to isolate backend servers, enforcing fine-grained access policies.
- Security Middleware: Performs deep inspection with threat detection and centralized logging.
- Backend MCP Servers: Simplified, isolated components focused solely on tool execution, leveraging the gateway for security.

## VI. Mapping to Security Frameworks

Before proceeding with the detailed threat mapping, it is helpful to briefly review the seven layers of the MAESTRO framework[5], which structures our analysis. MAESTRO breaks down the AI agent ecosystem into distinct layers: Foundation Models (Layer 1) provide core AI capabilities; Data Operations (Layer 2) and Agent Frameworks (Layer 3) manage data flow and tooling, respectively. Deployment Infrastructure (Layer 4) and Evaluation and Observability (Layer 5) cover the hosting environments and monitoring systems. Security and Compliance (Layer 6) spans across all layers, enforcing governance and control. Finally, the Agent Ecosystem (Layer 7) represents the external interface where business applications interact with users and third-party services. The table below provides a structured and comprehensive mapping between the MCP Gateway components and the MAESTRO framework. It serves as a valuable tool for securing enterprise AI integrations by systematically identifying and addressing potential threats.

TABLE I. MCP GATEWAY COMPONENT MAPPING TO THREATS

| MCP Gateway Area | MAESTRO Layer(s) | Key Threats | Primary Mitigations |
|---|---|---|---|
| Security Proxy | 4 (Deployment) | Denial of Service, Protocol Abuse | Rate limiting, Traffic shaping, Connection limits, Strict protocol validation, WAF rules |
| Auth Gateway | 3 (Agent Frameworks), 6 (Security & Compliance) | Agent Identity Attack, Lack of Auditability, Regulatory Non-Compliance | Strong OAuth 2.1 implementation, Scoped tokens, Secure key management, Logging, Privacy compliance adherence |
| Zero Trust Tunneling | 4 (Deployment), 6 (Security & Compliance) | Data Exfiltration, Side Channel Information Leakage | Network segmentation, Encryption in transit, Zero Trust Access, Service mesh, Immutable infrastructure, Security audits |
| Security Middleware | 5 (Evaluation & Observability) | Tool Poisoning, Threat Detection Failure | Threat detection (IDS), Content security policies, Input validation, Continuous behavior monitoring, Immutable audit trails |
| Backend MCP Servers | 3 (Agent Frameworks), 4 (Deployment) | Injection Attacks, Framework Evasion, Host System Compromise | Isolation, Strict input validation, Continuous framework hardening, Containerization, Host-based intrusion detection |

a. Threats as per Maestro Framework [5]

We further evaluated the security posture offered by the MCP Gateway architecture against the key mitigation strategies and identified threat categories, based on the security framework proposed by Narajala & Habler [4]. The MCP Gateway addresses the key security requirements outlined by Narajala & Habler by centralizing critical controls across network, application, identity, and monitoring layers. It enforces strong authentication and authorization through OAuth 2.1 integration, mitigating identity and access control subversion. Secure Zero Trust tunnelling protects communications from interception and lateral movement, aligning with their network segmentation principles. Application-level protections such as protocol validation, traffic inspection, and threat detection (e.g., tool poisoning mitigation through WAF and IDS integration) defend against resource exhaustion, tool misuse, and injection attacks. The gateway's continuous logging and anomaly detection capabilities support the operational security and continuous monitoring requirements emphasized in the framework. By consolidating these defences at a centralized ingress point, the MCP Gateway simplifies enterprise deployments while meeting the defence-in-depth, Zero Trust, and secure tool management practices advocated for robust MCP security.

## VII. Proof Of Concept Implementation

To validate the feasibility and security potential of the proposed MCP Gateway architecture, we developed a working proof of concept (PoC) using publicly available tools and minimal custom infrastructure. This implementation demonstrates how a modular, zero-trust gateway can securely mediate interactions between MCP clients and servers without embedding complex authorization logic into the backend.

The PoC was deployed on a hardened public facing Virtual Private Server (VPS) running Ubuntu Linux (Ubuntu 22.04 64 Bit). This setup provided a reliable and transparent environment for the deployment, ensuring compatibility with the components used in the implementation.

Pangolin[11] acted as the central management server, providing identity and access management. It securely exposed the local MCP server via WireGuard tunnels, preventing direct exposure to the public internet.

Pangolin leverages Traefik[12] for HTTP proxying, Let's Encrypt for Automated SSL certificates (https) and Gerbil for WireGuard tunnel management. Pangolin also incorporates dedicated internal components—including Badger for authentication services and Newt for WireGuard tunnel management - adopting a modular architecture that enhances security, simplifies maintenance, and ensures scalability. The following components were utilized with the specified versions for reproducibility:

TABLE II. SPECIFIC VERSIONS OF KEY COMPONENTS

| POC Software | Version |
| --- | --- |
| Traefik | V3.3.3 |
| Wireguard | 2023-12-11 release |
| Crowdsec | V1.6.8 |
| Crowdsec Bouncer Traefik plugin | V1.4.2 |
| Pangolin (inc Gerbil/Newt/Badger) | V1.2.0 |
| Docker | V28.1.1 |
| MCP Inspector | V0.10.2 |

We used the Anthropic MCP Inspector[7] as the primary client interface for initiating tool requests and inspecting protocol interactions. For external demonstration and testing, the Cloudflare AI Playground[6] was employed to simulate MCP interactions across the public Internet.

The system includes two stateless MCP servers, implemented with Server-Sent Events (SSE)—one hosted locally and one in the cloud—neither of which includes built-in authentication. These servers expose tools and resources accessible through the gateway.

A secure WireGuard-based tunnel connects the MCP Gateway to the local MCP server, with Pangolin serving as a tunnelled mesh reverse proxy. This setup ensures encrypted, identity-aware communication between the public gateway and internal services, preserving security even across distributed infrastructure.

The gateway itself is powered by Traefik, which functions as a secure proxy and ingress controller. Traefik applies several layers of middleware

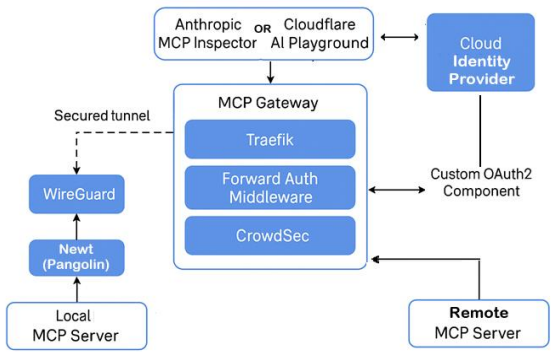

Fig. 3. Proof Of Concept Implementation

Forward Authentication Middleware delegates OAuth 2.1 flows to a custom-built component that interacts with identity providers (e.g., GitHub, Google) and returns authentication status and authentication token.

Intrusion Detection Middleware (CrowdSec) monitors traffic for anomalous behaviour and enforces behavioural bans [9]. This is achieved using Crowdsec Bouncer in a Traefik plugin[13] which provides virtual patching capabilities and WAF for advanced behaviour detection. Request Tracing and Logging are integrated for full observability and post-incident analysis. In addition, we developed a User Interface Manager to support onboarding and configuring new local MCP servers. This includes the ability to assign policies via Traefik middleware to MCP servers.

A core function of the MCP Gateway is its capability to abstract and manage the complexity of the OAuth authorization flow on behalf of MCP servers. This design enables developers to implement standard-compliant MCP servers using the Server-Sent Events (SSE) transport protocol, which can be securely deployed within enterprise networks. When an unauthenticated request is received, the gateway proxy[10] automatically issues a **401 Unauthorized** response, prompting the client to initiate a metadata discovery process. In response to the **.well-known/oauth-authorization-server** request, the MCP Gateway returns the necessary metadata, including the authorization server URI required for the OAuth exchange. Following successful user authentication, the client reattempts the original request, and the gateway proxy authorizes access based on the validated OAuth token, seamlessly enabling secure and authenticated communication with the MCP server.

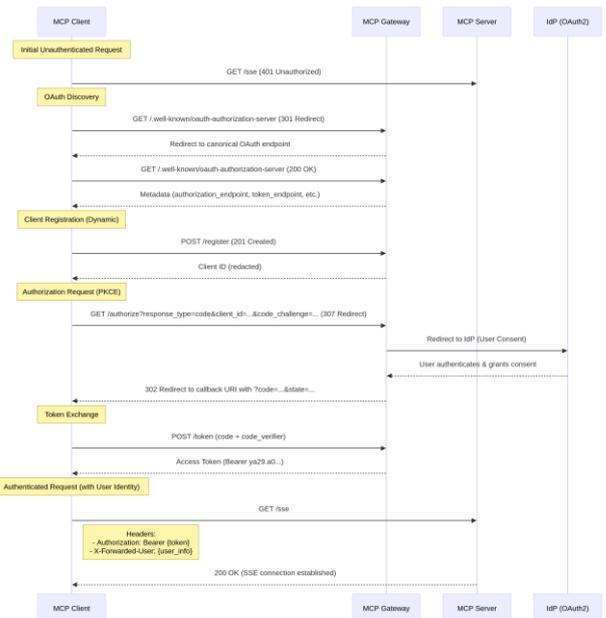

Fig. 4. External OAuth2 message flow

This PoC demonstrates that a dedicated MCP Gateway can effectively externalize authentication, authorization, and traffic inspection from the MCP servers, aligning with the updated OAuth 2.1 requirements of the official MCP Specification (2025-03-26). It further confirms that modern open-source components can be composed into a secure, flexible architecture suitable for enterprise-scale AI integration.

Fig. 5 below shows key snippets of the Traefik configuration demonstrating the implementation of essential security features. These include the setup of middleware for MCP authentication, OAuth metadata redirection, and CrowdSec protection.

```yaml
routers:
  helloworld-router:
    rule: "Host(`helloworld.yourdomain.com`)"
    entryPoints:
      - websecure
    middlewares:
      - mcp-auth@file      # Ensures traffic is authenticated via MCP auth
      - redirect-regex@file # Redirectsfor OAuth metadata
      - crowdsec@file      # CrowdSec bouncer to block bad actors
    service: helloworld-service
    tls:
      certResolver: letsencrypt  # HTTPS with Let's Encrypt certs
middlewares:
  mcp-auth:
    forwardAuth:
      address: "http://mcpauth:11000/sse"  # MCP auth service URL
      authResponseHeaders:
        - X-Forwarded-User   # Forward user information
  redirect-regex:
    redirectRegex:
      permanent: true
      regex: "^https://([a-z0-9-]+). yourdomain.com /.well-known/oauth
        -authorization-server"
      replacement: "https://oauth. yourdomain.com /.well-known/oauth
        -authorization-server"
  crowdsec:
    plugin:
      crowdsec:
        enabled: true
        crowdsecAppsecHost: "crowdsec:7422"  # CrowdSec appsec service
        crowdsecLapiHost: "crowdsec:8080"    # CrowdSec LAPI service
        crowdsecLapiKey: "your_api_key_here"
        captchaProvider: turnstile  # Captcha provider for protection
        httpTimeoutSeconds: 10
        updateIntervalSeconds: 15
        updateMaxFailure: 0
```

Fig. 5. Traefik Configuration – dynamic_config.yaml

## VIII. EVALUATION AND DISCUSSION

Qualitative testing showed the gateway's potential: The Oauth2 Gateway enforced authentication; Traefik blocked unauthenticated requests; invalid Authorization tokens were rejected. Traefik rate limiting and CrowdSec blocked excessive requests. Backend servers were isolated via WireGuard. Centralized authentication and WAF avoided replication on backend servers. Embedding security in each MCP server leads to duplication and complexity so offloading to a specific gateway shows merits.

Benefits:

The MCP Gateway decouples security from MCP servers, centralizes policy enforcement, enhances security posture through defence-in-depth and Zero Trust, and simplifies compliance with centralized logging.

Comparison with Alternatives:

Standard API Gateways lack MCP-specific threat understanding and Public MCP gateway solutions do not yet fully address enterprise self-hosted needs.

Limitations and Challenges:

Challenges and risks include the complexity of integrating components, performance overhead from security processes, managing keys/tokens, tuning threat detection rules, and ensuring reliable tunnel management. Another important consideration is the maturity of foundational components: Pangolin[11], which underpins the tunnelling and management framework, is a relatively new open-source project (less than six months old at the time of writing) and depends on several underlying technologies, including WireGuard. Over time, the security resilience of Pangolin and its ecosystem will become clearer through broader adoption, auditing, and community contributions.

Further Research:

Future work includes advanced AI/ML-based tool behaviour analysis, developing custom parsers, scenarios, and collections to enhance the MCP Gateway's intrusion detection capabilities, enabling more accurate detection of protocol-specific threats and contributing reusable security modules back to the broader open-source community, and more granular, context-aware authorization.

## IX. CONCLUSION

The proposed MCP Gateway architecture enables secure enterprise MCP adoption by centralizing security responsibilities. PoC results show feasibility in mitigating key risks. By abstracting complexity, it facilitates robust, secure, scalable, and spec-compliant AI integrations, crucial for trustworthy AI systems. While challenges remain, the gateway pattern offers a pragmatic path for managing MCP security at scale.